\documentclass{Interspeech}

\interspeechcameraready

\title{DiffSoundStream: Efficient Speech Tokenization via Diffusion Decoding}

\author[]{Yang}{Yang}
\author[]{Yunpeng}{Li}
\author[]{George}{Sung}
\author[]{Shao-Fu}{Shih}
\author[]{Craig}{Dooley}
\author[]{Alessio}{Centazzo}
\author[]{Ramanan}{Rajeswaran}

\affiliation{Google LLC.}{}{}
\email{\{yanghm, yunpeng, gsung, shaofu, dooleyc, centazzo, ramanan\}@google.com}
\keywords{speech tokenization, nerual speech codec, latent diffusion model}

\usepackage{comment}

\begin{document}

\maketitle

\begin{abstract}

Token-based language modeling is a prominent approach for speech generation, where tokens are obtained by quantizing features from self-supervised learning (SSL) models and extracting codes from neural speech codecs, generally referred to as semantic tokens and acoustic tokens. These tokens are often modeled autoregressively, with the inference speed being constrained by the token rate. In this work, we propose DiffSoundStream, a solution that improves the efficiency of speech tokenization in non-streaming scenarios through two techniques: (1) conditioning the neural codec on semantic tokens to minimize redundancy between semantic and acoustic tokens, and (2) leveraging latent diffusion models to synthesize high-quality waveforms from semantic and coarse-level acoustic tokens. Experiments show that at 50 tokens per second, DiffSoundStream achieves speech quality on par with a standard SoundStream model operating at twice the token rate. Additionally, we achieve step-size distillation using just four diffusion sampling steps with only a minor quality loss.

\end{abstract}

\section{Introduction}

Recent advances in self-supervised learning \cite{hubert,wav2vec2,wavlm, w2v-bert} and neural codecs \cite{encodec, soundstream} have allowed the transformation of speech signals into sequences of discrete tokens, effectively reframing speech modeling as a token-based language modeling problem. This paradigm shift has led to advances in a wide range of speech modeling applications such as speech continuation \cite{audiolm}, text-to-speech synthesis \cite{speartts, valle}, speech separation \cite{tokensplit}, speech translation \cite{polyvoice, s2st}, and end-to-end speech dialogue systems \cite{moshi}. A common thread across these solutions is their reliance on autoregressive (AR) modeling, applied either partially or fully to the token sequences.

A key challenge in AR-based speech modeling is balancing computational efficiency and reconstruction quality. The computational cost of AR models scales with the length of token sequences they process, making lower token rates, measured in tokens per second of speech, an attractive strategy for reducing inference latency, lowering computational demands, and simplifying the modeling task. Furthermore, a lower token rate extends the effective audio context length that an AR model can process given a fixed token context window. On the reverse side, aggressive token rate reduction risks degrading speech fidelity, necessitating careful optimization of the rate-distortion trade-off. This balance between efficiency and quality remains a central consideration in the design of token-based speech systems.

One way to mitigate distortion is by increasing the size of the codebook to enhance the information capacity per token. However, this approach is inherently unscalable, as the codebook size grows exponentially with the number of bits allocated to each token. In our work, we adhere to a fixed codebook size of 2048 per token, a standard configuration adopted in previous solutions \cite{rq, moshi}, and instead direct our efforts toward optimizing the token rate (tokens per second).

In this work, we limit our scope to non-causal/non-streaming tokenization during encoding and decoding, and stick to the hybrid semantic-acoustic tokenization approach \cite{audiolm,speartts,moshi, polyvoice, s2st}. Rather than examining specific token-based language modeling schemes, we focus solely on the tokenization process itself, investigating how to reduce the token rate for speech while preserving the perceptual quality of the reconstructed waveform.
\subsection{Contributions}

We propose DiffSoundStream, a tokenization model with three key contributions:

\noindent {\bf Semantic-conditioned acoustic tokenization}
By conditioning both the encoder and decoder of the SoundStream codec on semantic tokens, we explicitly reduce redundancy between semantic and acoustic token streams. Unlike previous hybrid approaches \cite{audiolm,speartts,moshi, polyvoice, s2st}, where semantic tokens merely guide the generation of acoustic tokens without participating in waveform decoding, our method encourages acoustic tokens to capture complementary information from semantic ones, allowing efficient reconstruction with fewer tokens.

\noindent {\bf Latent diffusion-based waveform decoding}
While standard neural speech codecs \cite{soundstream,encodec} employ deterministic GAN-trained decoders \cite{hifigan, melgan} without external stochasticity, we enhance the generative capability of the decoding process by training a latent diffusion model \cite{ddpm, diffusion_jascha} conditioned on semantic and coarse acoustic tokens. This approach overcomes the inherent quality ceiling of a deterministic decoding at low token rates, leveraging its iterative sampling process to synthesize perceptually critical waveform details.

\noindent {\bf Distilled 4-step diffusion sampling}
To mitigate the computational overhead of the generative process, we apply step-size distillation \cite{mmd} to reduce sampling iterations to just 4 steps with minor degradation in perceptual quality.

\subsection{Related works}
Previous works such as SpeechTokenizer \cite{speechtokenizer} and Mimi \cite{moshi} unify semantic and acoustic encoding via a single encoder by distilling information from an SSL. While distillation is necessary in transforming non-causal semantic information to a causal one, in our case we focus on non-causal tokenization and directly leverage SSL-derived semantic tokens without distillation. In addition to reducing token rate, hybrid AR and non-autoregressive (NAR) schemes \cite{valle,speechtokenizer, interleaved_token_depth, soundstorm} offer a complementary approach to improving the efficiency of token-based modeling and inference. We consider our scheme orthogonal to these designs.

\section{Modeling}

Our proposed DiffSoundStream model comprises three components, which we detail in the subsequent subsections.

\begin{itemize}
\item {\bf Semantic-conditioned SoundStream (SS-SC)}: A modified SoundStream autoencoder \cite{soundstream} where both the encoder and decoder are conditioned on semantic tokens derived from a pretrained WavLM model \cite{wavlm}. 

\item {\bf Continuous-latent SoundStream (SS-CL)}: A SoundStream \cite{soundstream} variant that replaces the discrete residual vector quantization (RVQ) bottleneck with a continuous latent space, trained using noise augmentation and regularization following the VAE formulation \cite{vae}. 

\item {\bf SoundStream latent diffuser}: A diffusion model trained to model the continuous latent distribution from {\bf SS-CL}, while conditioned on both the semantic tokens from WavLM and the acoustic tokens from {\bf SS-SC}. 
\end{itemize}

\subsection{Semantic-conditioned SoundStream model}\label{sec:ss-sc}

The SS-SC model we propose takes 24kHz audio waveform as input, and process it with four layers of encoder and decoder convolutional blocks that follow the architecture illustrated in Figure.~3 of \cite{soundstream}. The four blocks resample the temporal dimension at a factor of $[8, 8, 6, 5]$ (from waveform to bottleneck), resulting in a frame rate of $12.5$Hz at the bottleneck. The RVQ bottleneck employs 8 quantizers with 2048 codebook entries each, trained with uniform quantizer dropout to improve codebook utilization. We maintain the original SoundStream training recipes \cite{soundstream}, combining multi-scale STFT discriminators-based adversarial loss, feature matching loss, and reconstruction losses, as defined by Equation~(6) in \cite{soundstream}. Since we focus on non-causal tokenization, all the convolutions are non-causal with `same' padding.

Most token-based speech generation solutions adopt a hybrid tokenization scheme \cite{audiolm,speartts,moshi, polyvoice, s2st} that utilizes both semantic tokens derived from a self-supervised learning (SSL) model and acoustic tokens from a speech codec. Semantic tokens are often used as an intermediate modeling target to guide the generation of acoustic tokens and do not participate in the decoding of acoustic tokens to waveforms. To explore the information already embedded in the semantic tokens for acoustic tokenization, we propose to condition both the encoder and decoder of the SoundStream model with the semantic tokens through Feature-wise Linear Modulation (FiLM) layers \cite{film}. This conditioning encourages the SoundStream autoencoder to convey information complementary to that already captured in the semantic tokens and thus leads to a more efficient use of acoustic tokens.

\begin{figure}[t]
  \centering
  \includegraphics[width=\linewidth]{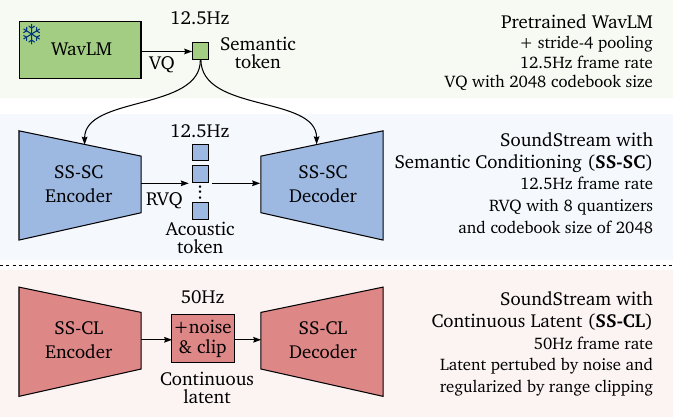}
  \caption{Overview of \textit{\textbf{SS-SC}} (top)
and \textit{\textbf{SS-CL}} (bottom).}
\label{fig:ae}
\end{figure}

As shown in Figure~\ref{fig:ae}, in SS-SC, we adopt WavLM to derive semantic information by following the same procedure in \cite{moshi} to downsample the feature from 50Hz to 12.5Hz with a non-causal average pooling at a kernel size of 8 and stride of 4. The downsampled features are quantized to 2048 centroids obtained by k-means clustering from a held-out dataset. This gives us a semantic token with token rate of 12.5Hz. 

The FiLM conditioning of semantic tokens is applied right before the last encoder block and right after the first decoder block. It first goes through a 256-dimensional embedding layer to convert to continuous features, and then followed by a transposed convolution to align with the temporal and depth dimensions of the corresponding activation that FiLM conditioning is applied upon.

\subsection{Latent diffusion model}

Diffusion model has emerged as a strong generative solution across  many domains. Here we exploit using latent diffusion \cite{ldm} to replace the GAN-trained deterministic SoundStream decoder to synthesize high-quality waveforms with reduced token rate. Specifically, we adopt the latent diffusion framework and rely on a continuous-latent SoundStream (SS-CL) model, shown in the lower half of Figure~\ref{fig:ae}, for latent encoding and decoding. We apply a diffusion network to model the distribution of such continuous latent conditioned on semantic and coarse acoustic tokens from WavLM and SS-SC. 

\subsubsection{Continuous-latent SoundStream model}
The configuration and the training recipe of the SS-CL network remains close to that of the original SoundStream model with the following changes: (1) The temporal striding factors of the four encoding/decoding blocks are set to [8, 5, 4, 3], leading to a frame rate of 50Hz, four times that of SS-SC. (2) The RVQ bottleneck quantization is replaced by a Gaussian posterior, effectively perturbing the continuous latent with Gaussian noise. Instead of applying KL loss term to regularize the latent distribution, we find that applying a simple range-clipping-based latent regularization works well enough in term of Gaussianizing the marginal distribution of each latent dimension. The latent dimension is set to 24; the noise standard deviation is set to be 0.2 of that of the empirical latent distribution; the noise addition is applied randomly for half of the training steps; and the latent range is clipped to be within $[-1, 1]$.

\subsubsection{WaveNet-based diffuser}

\begin{figure}[t]
  \centering
  \includegraphics[width=\linewidth]{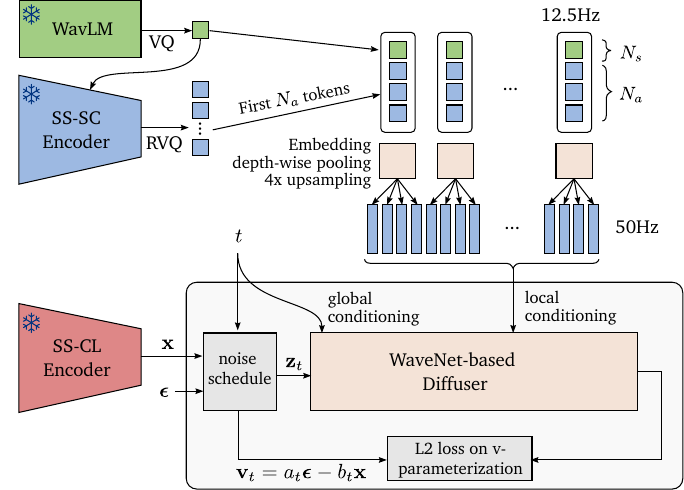}
  \caption{Training of the latent diffusion model that conditioned on the discrete semantic and acoustic tokens.}
\label{fig:diffusion}
\end{figure}

The latent representation obtained from the SS-CL model is normalized to unit standard deviation, which we denote as $\mathbf{x}$. We train a diffusion network to model the diffusion process, where the diffused latent $\mathbf{z}_t$ is defined as
\begin{align*}
\mathbf{z}_t = a_t \mathbf{x} + b_t \boldsymbol{\epsilon}
\end{align*}
Here $\boldsymbol{\epsilon}$ is a standard Gaussian vector with the same dimension as $\mathbf{x}$, and $t$ is sampled uniformly from $[0, 1]$. We adopt a modified cosine schedule \cite{iddpm}, defined by:
\begin{align*}
a_t = \cos\left(\theta_0 + (\theta_1 - \theta_0) t\right)\text{ and }
b_t = \sin\left(\theta_0 + (\theta_1 - \theta_0) t\right)
\end{align*}
where $\theta_0 = \tan^{-1}(e^{-3})$ and $\theta_1 = \tan^{-1}(e^{3})$. The values of $\theta_0$ and $\theta_1$ are chosen such that $t=0$ and $t=1$ yield latent signal-to-noise ratios (SNRs) of $60$dB and $-60$dB, respectively. Empirically, these SNRs ensure near-perfect waveform reconstruction and near-random noise after SS-CL decoding. The diffusion network follows $\mathbf{v}$-parameterization \cite{progressive_distillation, diffusion_elbo} and is trained with the following loss function without additional weighting:
\begin{align*}
L_\theta =& \mathbb{E}_{\mathbf{x}, \boldsymbol{\epsilon}, t}{\Big[}{\big|\big|}\text{Diffuser}_\theta(\mathbf{z}_t) - \left(a_t \boldsymbol{\epsilon} - b_t \mathbf{x} \right){\big|\big|}_2^2{\Big]}.
\end{align*}

The diffusion network utilizes a WaveNet architecture \cite{wavenet}, consisting of 5 non-causal WaveNet blocks, each with 5 residual layers employing dilation factors of [1, 2, 4, 8, 16]. The channel depth of each layer is set to 512, except for the input and output, which are tied to the latent dimension of the SS-CL model.

As depicted in Figure~\ref{fig:diffusion}, we condition the diffusion network on the semantic tokens from WavLM and the acoustic tokens from the semantic-conditioned SoundStream encoder. These discrete tokens are embedded and then average-pooled per frame. To address the frame rate discrepancy between the tokens and the continuous SS-CL latent, we apply nearest-neighbor upsampling with a factor of 4 to the pooled token embeddings. This aligns the conditioned token embeddings with the continuous latents along the temporal axis. The upsampled conditioning signal is injected into the WaveNet via local conditioning, while the time signal $t$ is incorporated as global conditioning.

\subsection{DiffSoundStream tokenization and decoding}

\begin{figure}[t]
  \centering
  \includegraphics[width=\linewidth]{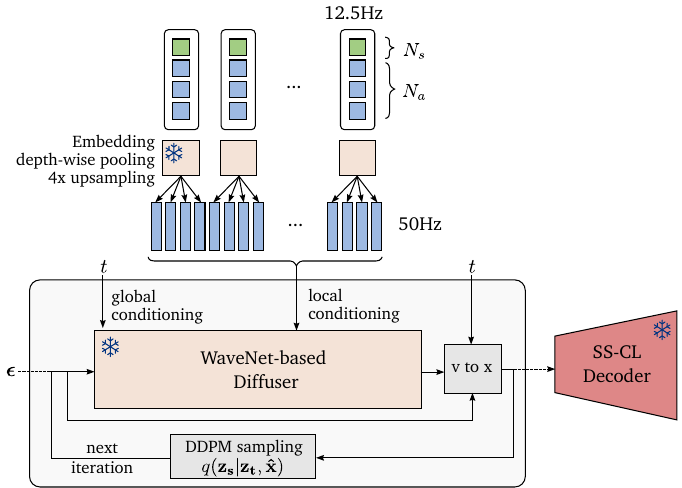}
  \caption{Token to waveform decoding.}
\label{fig:token-to-wav}
\end{figure}

DiffSoundStream integrates WavLM and SS-SC-based tokenization, with latent diffusion-driven decoding. We outline the tokenization and decoding workflow below, as visualized in Figure~\ref{fig:diffusion} and Figure~\ref{fig:token-to-wav}.

\noindent \textbf{Tokenization}  Input speech is first encoded into a single-level semantic tokens using WavLM features temporally-pooled and quantized via k-means clustering (2048 codebook entries). The SS-SC encoder, conditioned on the semantic tokens, produces 8 RVQ acoustic tokens per frame. In total, each frame contains 9 tokens at a frame rate of 12.5Hz, translating to a maximum token rate of 112.5 tokens/second (9 tokens/frame × 12.5 frames/sec) with a bitrate of 1.2375 kbps (9 tokens × 11 bits/token × 12.5 frames/sec). At decoding time, $N_s$ semantic tokens (0 or 1) and $N_a$ acoustic tokens (1-8) are used in the diffusion decoder.

\noindent {\bf Decoding} The combined $N_s+N_a$ semantic and acoustic tokens, after going through depthwise-pooling and temporal-upsampling, are taken as the conditioning source of the latent diffusion decoder. DDPM \cite{ddpm} sampling is applied to generate speech latent, which is then decoded into speech waveform by the SS-CL decoder.

Note that the SS-SL decoder and SS-CL encoder are only relevant during training. Neither of them participates in the tokenization or decoding of DiffSoundStream.

\subsubsection{Step-size distillation}

While diffusion models typically require extensive sampling steps (100 iterations in our baseline), we apply moment matching distillation (Algorithm~2 in \cite{mmd}) to dramatically reduce computational overhead. This technique trains a student model that approximates the posterior distribution of DDPM sampling of the original teacher diffusion network by optimizing a reverse-KL objective. The gradient of this reverse-KL objective is approximated with the help of an auxiliary diffusion model \cite{dmd2}. Experiments show that we can compress the 100-step sampling processing into 4-steps with a minor loss in quality. The distillation preserves the conditioning mechanism, enabling efficient synthesis without architectural modifications.

\section{Experiments}

\begin{figure}[t]
  \centering
  \includegraphics[width=1\linewidth]{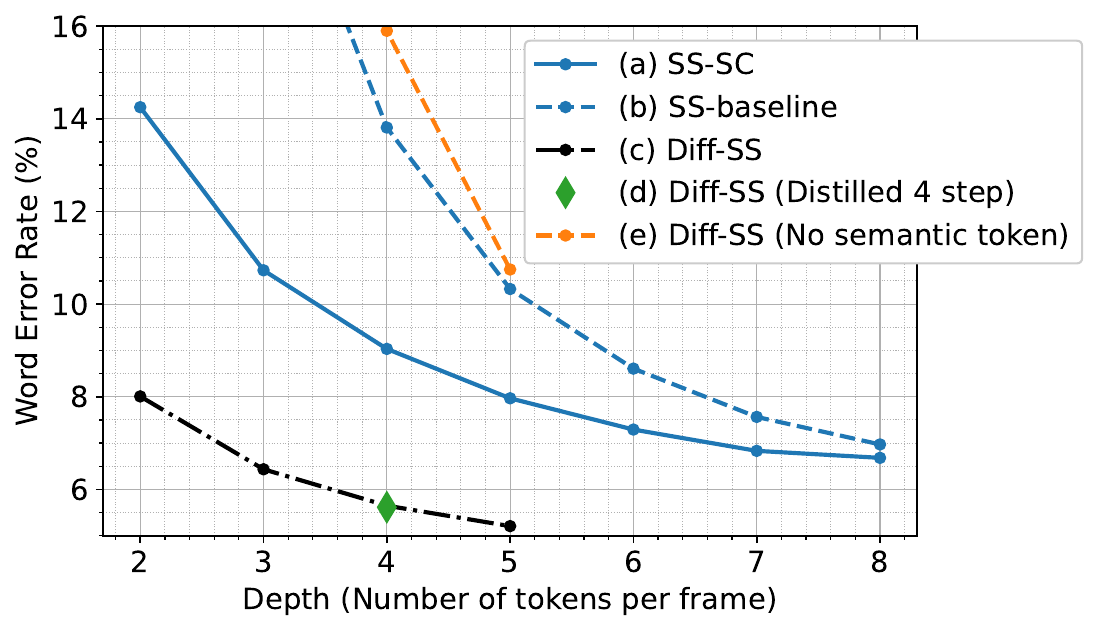}
  \caption{Word-error-rate on LibriTTS-test-clean comparing the baseline SoundStream (SS-Baseline), semantic-conditioned SoundStream (SS-SC), and DiffSoundStream (Diff-SS) as a function of tokens/frame at 12.5Hz frame rate.}
\label{fig:wer}
\end{figure}

\begin{figure}[t]
  \centering
  \includegraphics[width=0.85\linewidth]{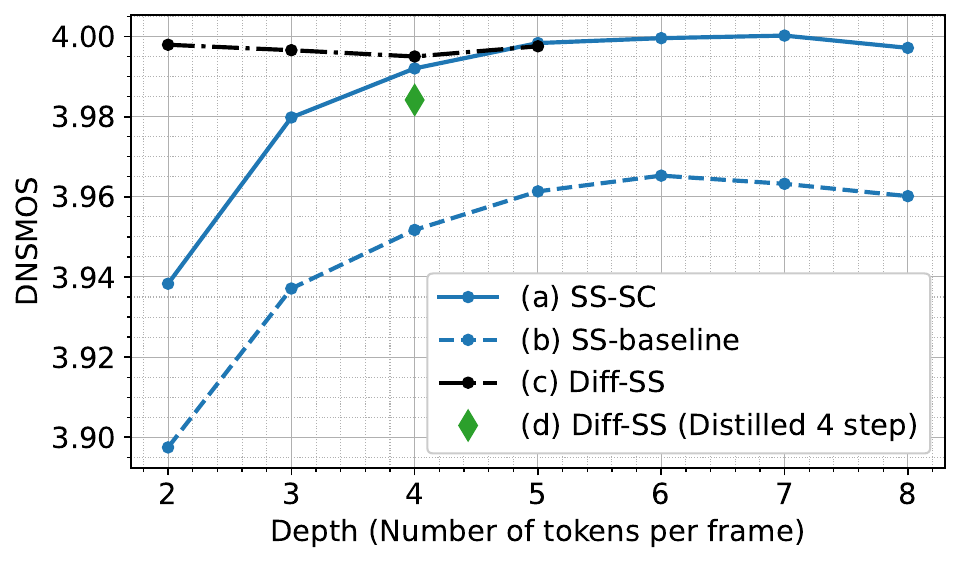}
  \caption{DNSMOS study.}
\label{fig:dnsmos}
\end{figure}

The three components SS-SC, SS-CL, and the diffusion model are all trained with a private dataset containing more than 10k hours of English conversational speech using the Adam optimizer \cite{adam}. SS-SC and SS-CL are optimized with $1\mathrm{e}{-4}$ learning rate for 1 million steps. The diffusion model is trained for 4 million steps with a learning rate schedule that decays from $3\mathrm{e}{-4}$ to $1\mathrm{e}{-5}$. The 4-step distilled diffusion network, together with the auxiliary network used for estimating the gradient of the distillation training, is initialized from the teacher version and finetuned for 150k steps at a learning rate of $1\mathrm{e}{-6}$ with $\beta_1$ (exponential decay rate of the first moment of gradient) set to $0$.

We evaluate our solution on Libri-TTS \cite{libritts} test-clean subset by examining the intelligibility and perceptual quality of the reconstructed speech after tokenization and decoding. Specifically, we measure the word-error-rate (WER) of the reconstructed speech transcribed with greedy decoding of a HuBERT-Large-based ASR model\footnote{\href{https://huggingface.co/facebook/hubert-large-ls960-ft}{https://huggingface.co/facebook/hubert-large-ls960-ft}} without any language models. For perceptual quality evaluation we rely on DNSMOS \cite{dnsmos}, an objective measure, and MUSHRA \cite{mushra}, a subjective listening test.

In Figure~\ref{fig:wer}, we show the WER of the following five solutions as a function of tokens/second (a.k.a. token depth). In Figure~\ref{fig:dnsmos}, we further visualize the DNSMOS score of the first four solutions.
\begin{itemize}[leftmargin=.2in]
    \item[(a)] SS-SC: the semantic-conditioned SoundStream model as described in Section~\ref{sec:ss-sc}. The figure shows the WER of SS-SC auto-encoding using 1 semantic token and 1-7 acoustic tokens (so the token depth ranges from 2 to 8).
    \item[(b)] SS-Baseline: a standard SoundStream \cite{soundstream} model with the same configuration as (a) except that no semantic token is applied. The token depth reflects the number of acoustic tokens applied for auto-encoding.
    \item[(c)] Diff-SS: the proposed DiffSoundStream model that combines SS-SC encoding with diffusion decoding. The plot shows the WER from four diffusion models trained with $N_s=1$ and $N_a=1,2,3,4$ (so the token depth $N_s+N_a$ range from 2 to 5).
    \item[(d)] Diff-SS (Distilled 4-step): The step-size distilled version of a Diff-SS model trained with $N_s=1$ and $N_a=3$.
    \item[(e)] Diff-SS (No semantic token): An ablated version of (c) where the diffusion model is trained by conditioning 2 to 5 acoustic tokens from SS-Baseline.
\end{itemize}

The comparison of (a) and (b) demonstrates a clear advantage of applying semantic token as conditioning for the SoundStream autoencoder: SS-SC is uniformly better than SS-Baseline in WER across token depths and the gap widens as the depth reduces. There is a consistent improvement in DNSMOS as well. Comparing (b) and (c) we can see that the diffusion decoder further boosts both the intelligibility and perceptual quality of the decoded speech. It is  surprising to see that the DiffSoundStream in (c) achieves better WER comparing to SS-SC (a) under at the same token depth considering that they share the same encoder. This suggests that the WavLM semantic-tokens plus the SS-SC acoustic tokens capture more intelligibility information that its GAN-trained decoder is able to utilize. Furthermore, it is interesting to observe that the DNSMOS score of (c) stay roughly constant as a function of token depth, which is a testament to diffusion model's strong generation capability in filling in missing acoustic details even when there is a lack of acoustic details at low token rate. The WER gap between (c) and (e) shows that, at low token rate, it is crucial for the diffusion model to condition on semantic tokens to be able to generate speech with consistent content. Looking at (d), we can see that at depth of 4 ($N_s=1, N_a=3$), the 4-step distilled version of the diffusion decoder yields the same WER with a slightly degraded DNSMOS score. 

\begin{figure}[t]
  \centering
  \includegraphics[width=0.85\linewidth]{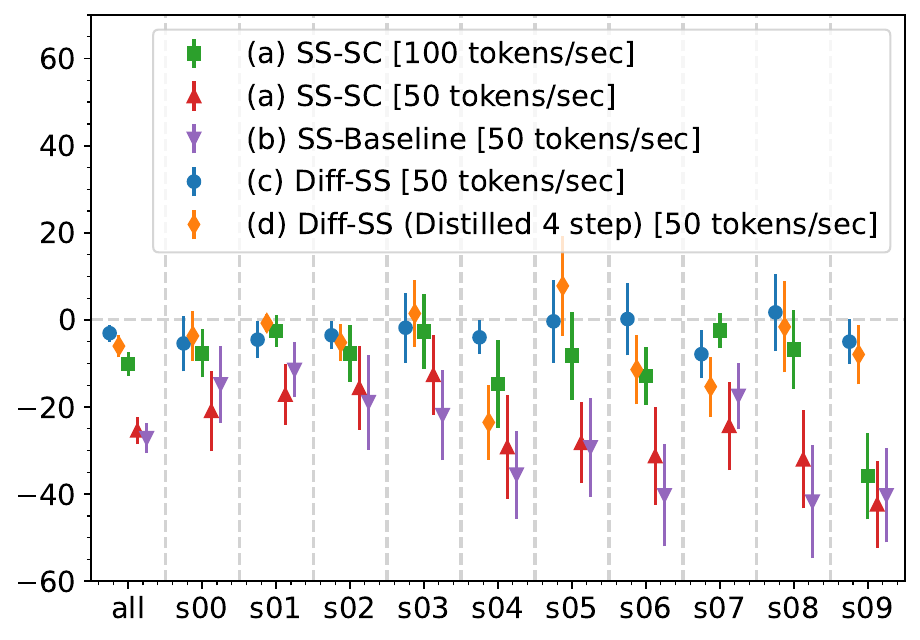}
  \caption{Average MUSHRA score differences relative to clean audio, together with 95\% confidence intervals. First column is the mean of each model over all audio clips. Number in the parenthesis indicate token rate in tokens/second.}
\label{fig:mushra}
\end{figure}

We further supplement the objective evaluation with a MUSHRA subjective listening test\footnote{The 10 audio clips are sampled uniformly from LibriTTS test-clean omitting those less than 5s.
}, the results of which is shown in Figure~\ref{fig:mushra}. We zoom in on the case when token depth is 4, or equivalently 50 tokens/second at 12.5Hz frame rate, for (a)(b)(c)(d), and compare them with the 100 token rate result from (a). It is evident that both DiffSoundStream and the 4-step distilled version at 50 tokens/second achieves overall better quality compared to SS-SC at twice the token rate.

\section{Conclusions and limitations}

We propose DiffSoundStream, a speech tokenization solution that derives acoustic tokens conditioned on WavLM semantic tokens and utilizes latent diffusion for efficient speech decoding. There are two major limitations: it only supports non-causal tokenization and decoding, and the semantic information is limited to a single quantization level. A future direction could be to extend it to causal tokenization and decoding \cite{ar_diffusion} and study the general trade-off of semantic and acoustic token allocation.

\bibliographystyle{IEEEtran}
\bibliography{mybib}

\end{document}